\documentclass[showkeys,
reprint,
aps,
pre
]{revtex4-2}
\usepackage{graphicx}
\usepackage{xurl}

\begin{document}

	\title{On Spherical Shock Wave focusing in Air - a Computational Study}
	
	\author{Saranyamol V. S.}
	\author{Soumya Ranjan Nanda}%
	
	\author{Mohammed Ibrahim S.}
	\email{ibrahim@iitk.ac.in}
	\affiliation{Hypersonic Experimental Aerodynamics Laboratory\\Department of Aerospace Engineering\\
		Indian Institute of Technology Kanpur, Uttar Pradesh, India. 208016.
	}

	\begin{abstract}
		A detailed numerical study on the phenomenon of Shock Wave focusing in air is carried out. The focusing phenomenon is achieved with the help of a shock tube and a converging section attached to it. The planar shock generated inside the shock tube is converted to spherical shock with the help of the converging section and is focused to a point. High-temperature effects like temperature-dependent $C_p$ variation and chemical reactions corresponding to dissociated air are included in the simulation. The chemical reactions including the dissociation, recombination and ionization of nine species of air including three ions ($N_2$, $O_2$, N, O, NO, Ar, $NO^+$, $O^+$ and $Ar^+$) are monitored throughout the simulation. The effect of driven section filling conditions such as initial pressure and temperature on focusing parameters is studied. The variation in the initial fill temperature is found to affect the flow properties much more as compared to the change in initial fill pressure, while maintaining the same shock strength. The effect of incident shock strength on shock wave focusing is also investigated. It is observed that as the strength of the shock increases, the conditions like temperature and pressure at the focusing point increases and thereby increasing the reaction rate of all the reactions. 
	\end{abstract}
	\maketitle
	
	\pagebreak
	
	\section{\label{sec:level1}Introduction}
	
	Shock Waves (SW) are disturbances that travels faster than the local acoustic speed in a medium and usually formed as an outcome of impulsive energy discharge. SW has the ability to rapidly dissipate the inherent energy, thereby causing drastic change in fluid properties. This property of SW has enabled researchers to explore and investigate the possibility of SW focusing; a method where SW is confined or focused to a small region in space. The phenomenon of focusing a SW to a point/region leads to very high energy concentration and typically finds application in the field of medicine (SW lithotripsy), material science \cite{glass1976} and nuclear technology \cite{glass1982}. Most of the times, the temperature in the focused region is so high that the enclosed gas starts radiating \cite{veronica2007} which can be depicted during supernova collapse \cite{arnett1989} and gas bubble sonoluminescence \cite{kondic1995}. Applications of SW focusing phenomenon in the aerospace industry is enormous. Shock focusing ignition techniques enable in increasing the efficiency of a pulse detonation engine \cite{zhang2016}. Flow conditions that exhibit within in the shock layer of the spacecraft as it encounters various planetary entry/re-entry conditions can also be simulated through shock focusing. Detailed investigation to generate this radiating equilibrium and non-equilibrium flowfield using ground based facilities is scarce. In past, few methods were employed to focus a SW \cite{apazidis2019}, which includes shock tube exploration, exploding wire technique, usage of micro explosives etc.
	
	Several researchers have attempted to attain SW focusing through analytical, numerical and experimental approaches out of which Guderly (1942) \cite{guderley1942} was the first to investigate the convergence of SW analytically. A self-similar analytical solution was proposed for the radius of converging SW as a function of time. Simultaneously, experimental methods for achieving SW focusing was developed which encompasses tear-drop insert inside a shock tube \cite{perry1951}, hemispherical implosion chamber \cite{roig1977}, parabolic reflector inside shock tube \cite{sturtevant1976}, annular shock tube \cite{veronica2007} etc. Most of the SW focusing achieved through shock tube was performed for a rectangular cross-sectional geometry which thereby resulted in 2-dimensional cylindrical SW. The exception being the facility at the KTH Royal Institute of Technology, Sweden, where focusing of a spherical SW was achieved with the help of a shock tube having circular cross-section \cite{maltePhd}\cite{liverts2016}\cite{sembian2020}. It was also reported that in the case of spherical SW focusing, the energy concentration in the core region is higher as compared to cylindrical way of focusing the shock wave \cite{mine}.
	
	During the spherical SW focusing, a smooth curvature transformation section is attached to the driven section end, upon entering which the planar SW gets smoothly transformed to spherical shape with minimal diffraction. After this SW implosion, very high temperature region (~27000 K in Argon test gas) was observed at the instance of focusing \cite{malte2012}. Also, a review of literature shows that spherical SW focusing and characterizing the flow field properties with air as test gas is minimal. 
	
	With this backdrop, the current work focuses towards simulating spherical SW focusing in air medium through numerical approach to have a better understanding of the flow field and its properties. The computational domain for this study involves a shock tube along with an attached converging section which enables in generating spherical shock with minimum diffusion. With shock wave focusing to a point, rapid amplification of energy concentration could trigger dissociation, ionization and recombination of constitutive molecules as well as ions. The major components of atmospheric air being oxygen, nitrogen, and argon, the variation in concentration of these species along with ions are monitored during SW focusing. Moreover, the influence of the change in initial fill conditions, such as fill temperature and fill pressure, as well as the initial shock strength on the focusing phenomenon are also carried out. Quantitative and qualitative assessment of alternation in flow parameters at the focused region and temporal variation of different species concentration are performed for estimating the key parameter which affects the focusing phenomenon to a larger extent.

	\section{\label{sec:level2}Numerical Modeling and Approach}
	
	The current numerical simulations are carried out with commercially available CFD software ANSYS Fluent V-18.2. An initial assessment of the flow was obtained by performing perfect gas simulations. Later, in view of the probable dissociation of air molecules, high-temperature effects along with reaction models are incorporated into the simulations.
	
	\subsection{\label{sec:level2.1}Geometry Details}
	
	The geometry for the present simulation includes a shock tube having smoothly converging test section at the end in order to achieve SW focusing, and the dimensions are based on the shock tube test facility present in the Hypersonic Experimental Aerodynamic Laboratory (HEAL), Department of Aerospace Engineering, IIT Kanpur. The facility with 85mm internal diameter has a 1 m driver tube and 7 m long driven section with a 296 mm converging portion attached to its end. The exit diameter of the convergent section is measured to be 0.6 mm. Moreover, in order to have comparative assessment with experimental solution, 26 mm from the end of the converging part is designed as a conical section having 21$^{\circ}$ semi apex angle. The profile for the contoured converging test section is generated as per the geometric relations proposed by Malte \cite{maltePhd}, with an intention to have minimum diffusion of the moving shock.  
	
\begin{figure}[h!]
	\centering
	\includegraphics[width=0.9\linewidth]{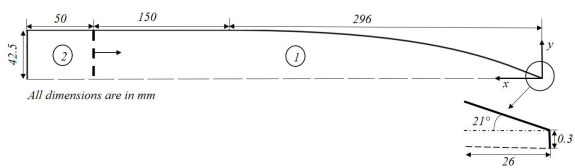}
	\caption{Schematic diagram of the flow domain of the converging section attached to the shock tube end. The 26 mm long conical section is shown as an insert}
	\label{domain}
\end{figure}

	Furthermore, it is computationally expensive to simulate the entire shock tube with attached converging section. Therefore, keeping in view of the region of interest, i.e. near focusing region, the computational domain is restricted till 200mm of the constant area section as shown in Figure ~\ref{domain}. This length is finalized in conjunction with the x-t diagram \cite{anderson2000} for the corresponding test cases with an intention to avoid any flow interaction inside the test domain during the test duration.
	
	Further reduction in the computational complexity and cost is achieved by considering half of the domain due to the axis-symmetric nature of the problem. The final computational domain as shown in Figure ~\ref{domain} is divided into two segments for initialization purpose where the region 1 is assigned with driven gas filling parameters. However, to replicate a planar moving shock of strength of Ms inside the domain, Rankine- Hugoniot relation is used to obtain the corresponding flow properties associated with the incident shock Ms and are given as the initial condition for region 2.
	
	\begin{figure}
		\centering
		\includegraphics[width=0.9\linewidth]{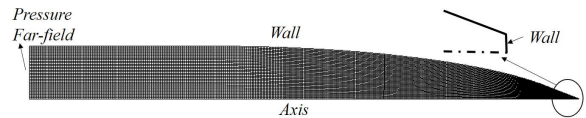}
		\caption{Mesh and boundary condition of the computational domain}
		\label{mesh}
	\end{figure}
	
	The boundary conditions associated with the domain is shown in Figure ~\ref{mesh} where the base of the domain is provided with axis boundary condition to imitate the axis-symmetric simulations. The extreme left edge of the domain is considered as pressure far-field with properties same as assigned to zone-2. This ensures a continuous influx of the shock-induced test gas into the domain. The top and right boundary is assigned with no-slip wall condition. 
	
	\subsection{\label{sec:level2.2}Numerical approach}
	
	The simulations are carried out using commercially available software ANSYS Fluent which incorporates finite volume approach to solve the governing equations. A density-based, transient-implicit solver is used to solve continuity, momentum and energy equation. AUSM flux scheme and second-order spatial discretization are adopted as solution methods to improve solution accuracy. Air with initial concentration of 76\% Nitrogen, 23\% Oxygen, 1\% Argon is considered as test gas for most of the simulations. Initially, inviscid perfect gas simulations are carried out to have information on the qualitative aspect of the flow structures evolved during focusing along with magnitudes of different flow parameters. These simulations are also extended to ensure necessary validation with experimental test cases. It is observed that at the instance of focusing, the temperature magnitude shoots up rapidly to very high values ranging from 4000 K to 6000 K which can trigger reaction between constitutive elements of air molecule. To precisely model these reactions,  simulations with included high temperature effects which encompass temperature-dependent Cp variation and inclusion of chemical reactions are carried out.

	\begin{table}[b]
		\caption{\label{tab:table1}%
			Reactions included for simulations
		}
		\begin{ruledtabular}
			\begin{tabular}{lcr}
				\textrm{No.}&
				\textrm{Forward Reactions}&
				\textrm{Rate Constant (cm3/mol s)}\\
				\colrule
				1 & $O_2$ +M $\rightarrow$2O+M  & 3.6$\times10^{18} T^{-1.0} e^{-5.95 \times 10^4/T}$\\
				2 & $N_2$+M$\rightarrow$2N+M(O) & 1.9$\times10^{17} T^{-0.5} e^{-1.13\times 10^5/T}$\\
				3 & NO+M$\rightarrow$ N+O+M($O_2$)  & 3.9$\times10^{20} T^{-1.5} e^{-7.55\times10^4/T}$\\
				4 & $N_2$+O$\rightarrow$ NO+N  & 3.2$\times10^9 T^1 e^{-1.97\times10^4/T}$\\
				5 & NO+O$\rightarrow$ N+$O_2$  & 7.0$\times10^{13} e^{-3.8\times10^4 /T}$\\
				6 & $N_2$+N$\rightarrow$2N  & 4.085$\times10^{22} T^{-1.5} e^{-1.13\times10^5/T}$\\
				7 & $O_2$+O$\rightarrow$2O+O  & 9.0$\times10^{19} T^{-1.0} e^{-5.95\times10^4/T}$\\
				8 & $O_2$+$O_2$$\rightarrow$2O+$O_2$  & 3.24$\times10^{19} T^{-1.0} e^{-5.95\times10^4/T}$\\
				9 & $O_2$+$N_2$$\rightarrow$2O+$N_2$  & 7.2$\times10^{18} T^{-1.0} e^{-5.95\times10^4/T}$\\
				10 & $N_2$+$N_2$$\rightarrow$2N+$N_2$  & 4.7$\times10^{18} T^{-0.5} e^{-1.13\times10^5/T}$\\
				11 & NO+M$\rightarrow$ N+O+M(O)  & 7.8$\times10^{22} T^(-1.5) e^{-7.55\times10^4/T}$\\
				12 & N+O$\rightarrow$ $NO^+$+$e^-$  & 5.3$\times10^{12} e^{-3.1\times10^4/T}$\\
				13 & O+ $e^-$$\rightarrow$ $O^+$+2$e^-$  & 3.9$\times10^{33} T^{-3.78} e^{-1.5855\times10^5/T}$\\
				14 & Ar+ $e^-$$\rightarrow$ $Ar^+$+$2e^-$  & 5.7$\times10^8 T^{-1.5} e^{-1.35\times10^5/T}$\\
			\end{tabular}
		\end{ruledtabular}
	\end{table}

	$C_p$ is assumed to be piecewise polynomial function of temperature \cite{cp} and mixing law formulation is used for calculating the effective $C_p$ of the test gas. Similarly, reaction models including 11 dissociation and recombination reactions \cite{desai2016} along with 3 ionization reactions\cite{shoev2017}\cite{park1989} are taken into account for estimating variation in species concentration of Nitrogen ($N_2$), Oxygen ($O_2$), Nitric Oxide (NO), Atomic Nitrogen (N), Atomic Oxygen (O) and Argon (Ar) as well as ions like $NO^+$, $O^+$ and $Ar^+$. Details of the reactions used in the current simulation are listed in Table~\ref{tab:table1}. To estimate these species concentrations, species transport equation is solved where finite rate model is opted for computing the rate of creation or destruction of a species. Nevertheless, for a specific reaction, the forward reaction rate is effectively estimated through the use of Arrhenius equation (Equation~\ref{1}) with constant parameters obtained from previous studies \cite{desai2016}.

	\begin{equation}
		\label{1}
		K_f = AT^ {(\beta)} e ^{(E/RT)}
	\end{equation}

	Where, $K_f$: Forward rate constant, A: Pre-exponential factor, $\beta$: Temperature exponent, E: Activation energy for reaction and R: Universal gas constant.

	\begin{table}[t]
		\caption{\label{tab:table2}%
			Test cases for the current simulations
		}
		\begin{ruledtabular}
			\begin{tabular}{lcr}
				\textrm{Test-Cases}&
				\textrm{P1 (Pa)}&
				\textrm{T1 (K)}\\
				\colrule
				Case-1: M1.5 & 101325  & 293\\
				Case-2: M1.5-P30 & 131723 & 293\\
				Case-3: M1.5-T30 & 101325  & 380\\
				Case-4: M2.05 & 101325  & 293\\
			\end{tabular}
		\end{ruledtabular}
	\end{table}
	
	Table~\ref{tab:table2} lists the initial conditions with the flow that is being considered for the current simulations. The prime intent behind the variations in initial conditions is attributed to estimate the effect of different filling condition towards shock focusing phenomenon in terms of assessing the magnitude of different flow parameters and species concentration. Case-1 is taken as a base case study, where the incident shock strength Ms is fixed to be 1.5, and the driven gas conditions are set as atmospheric. The nomenclature for case-1 is designated   as ‘M1.5’. To obtain the effect of driven gas filling pressure on the shock focusing phenomenon, Case-2 is generated where the initial pressure in the region 1 ($P_1$) is increased by 30\% keeping the incident shock strength constant (M1.5-P30). Similarly, the effect of initial filling temperature is estimated through Case-3 in which only the driven gas viz. zone-1 temperature ($T_1$) is increased by 30\% (M1.5-T30). Furthermore, the induced effect of increase in the shock strength on focusing phenomenon is obtained by varying the incident shock Mach number Ms to 2.05 as presented in Case-4. 
	
	During the simulation for aforementioned test cases, the temporal variation of static temperature, pressure and mass fraction of different species is monitored at the focusing point. Besides, the static pressure at several points along the axis of the domain is also monitored which enabled to obtain the change in shock Mach number as the planar shock enters into the converging section. 
	
	Prior to actual simulations, considering the transient nature of the shock tube problem, it is hereby planned to carry out grid as well as time step independence study. In line with this, inviscid perfect gas simulation is performed with initial conditions corresponding to Case-4 where the incident shock strength is 2.05.
	
	\subsubsection{Grid independency test}
	
\begin{figure}[h!]
		\centering
		\includegraphics[width=0.7\linewidth]{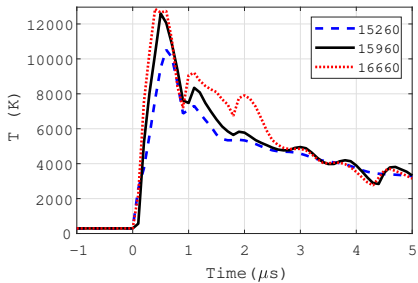}
		\caption{Grid independency test showing the peak temperature for perfect gas simulations of case M2.05 at the focusing point}
		\label{grid}
	\end{figure}

	Meshing for the present computation, which can be seen in Figure ~\ref{mesh}, was carried out by quadrilateral mapped facing method with bias factor given to the zone of interest viz. converging section. Grid independency test was carried out in two steps. Primarily, meshing parameters are changed in the constant area section as well as in the initial part of the converging section to ensure that the flow properties are not changing. Further, the grid concentration inside the conical section (refer Figure ~\ref{domain}) being the decisive parameter in determining the focusing phenomenon more accurately, it is thereby varied and illustrated in Figure  ~\ref{grid}. The variation of the peak temperature obtained at the focusing point for different grid sizes inside the conical section shows minute change for 15960 and 16660 elements. Therefore, for future simulations total number of elements is considered to be 15960.
		
	\subsubsection{Timestep independency test}
	
\begin{figure} [h!]
	\centering
	\includegraphics[width=0.7\linewidth]{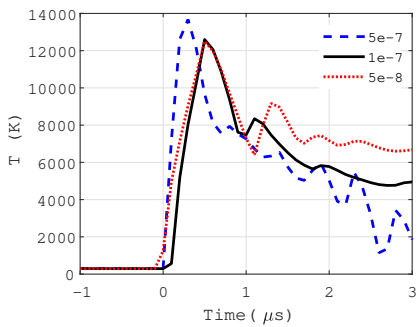}
	\caption{Timestep independency test peak temperature for perfect gas simulations of case M2.05 at the focusing point}
	\label{timestep}
\end{figure}

	The flow duration inside the shock tube is of the order of few milliseconds; however, as it can be seen from Figure ~\ref{grid}, the peak temperature inside the domain happens to be for few microseconds only. Henceforth, the timestep of the simulation also plays a vital role in deducing the proper estimate for the flow properties. Independency test for timestep is also ensured through detailed simulations for the initial filling condition associated to Case-4. The result for the peak temperature at the focusing point for different time step is shown in Figure ~\ref{timestep}. 

	It can be observed that for smaller time steps, the peak temperature magnitude is almost equal with some discrepancy in trend. Minute variation in temperature profile with decrement in time step can be ascertained as the Courant number approaches to unity. Nevertheless, to reduce the computational cost, 1$e^-7$ is chosen as the suitable time step for further simulations.
	
	\subsection{\label{sec:level2.3}Validation of results}	
\begin{figure}[h]
		\centering
		\includegraphics[width=0.7\linewidth]{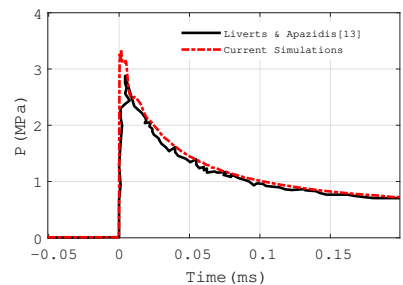}
		\caption{Validation of the current simulations with experimental results showing the peak pressure obtained at the focusing region }
		\label{vlidation}
	\end{figure}

	After obtaining the optimal meshing and simulation parameters, to establish the correctness of the numerically obtained flow parameters, it is thereby planned to validate results with experimental outcomes. In view of this, simulations are carried out for shock Mach number of 3.25 with argon as test gas \cite{liverts2016}. Argon being a monoatomic and inert gas, inviscid simulations with perfect gas assumption is performed with driven section filling pressure of 10 kPa. Figure ~\ref{vlidation} shows the comparison between experimentally and numerically obtained temporal variations of pressure at the focusing point. The pressure variation seems to have a good agreement with the experimental result in terms of trend. However, peak magnitude for the numerical simulation has a variation of 12.5\% as compared to the experimental result. Nevertheless, the experimental uncertainty in pressure measurement is also reported to be 12\% \cite{sembian2020}. The close match of pressure signal thereby provides confidence for the flow parameters that is going to be achieved during the actual test cases.
	
	\section{\label{sec:level3}Results and Discussion}
	
	After successful validation, simulations are carried out towards comprehensive investigation of analyzing the behavior of shock within the converging section in terms of variation of flow properties and species concentration. 
	
	\begin{figure}[b]
		\centering
		\includegraphics[width=0.9\linewidth]{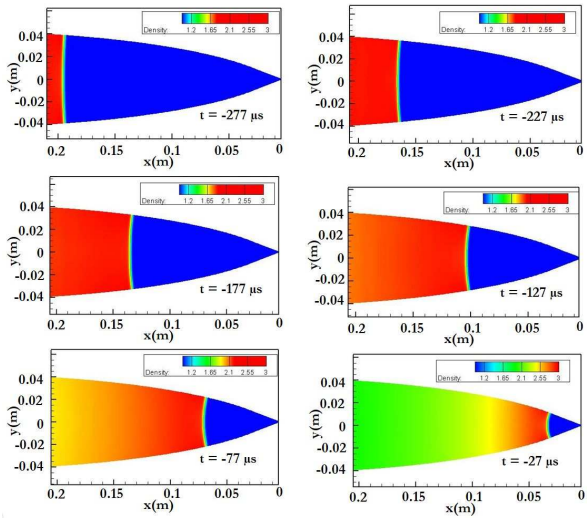}
		\caption{Shock trajectory inside the converging section depicted with the help of the density contour at various instances from a time interval of 277 $\mu$s time to 27 $\mu$s before the shock reaches the focusing point}
		\label{densityBig}
	\end{figure}
	
	Initially, perfect gas simulation for Case-1 is performed to get an idea on the focusing phenomenon. The variation in the shape of the shock while passing through the converging section is investigated through detailed analysis. Change of shock shape from planar to spherical through gradual curving can be seen from Figure ~\ref{densityBig}. This transformation of shape mostly relies on the shape of the converging section, as the shock curving initiates at the associated wall. Additionally, as the geometry of the converging section is designed to have less diffusion of the shock, thus it makes the SW foot to remain perpendicular to the wall and attains spherical shape with minimum losses to the shock front.
	
	Six instances of density contour prior to shock arrival at the focusing point are portrayed in Figure ~\ref{densityBig}. The axis-symmetric domain is mirrored to obtain a 2D planar view. The time instance at which the shock reaches the focusing point is taken as reference for all the contour plots. At time 277$\mu$s before focusing, the shock is clearly depicted to be a planar shock. However, along with marching in time, the shock approaches towards the focusing point and the shape of the shock curves gradually. At -27$\mu$s, spherical curvature of the shock is evident. Besides, it is noticed that the shock has taken 440$\mu$s to travel the entire converging section; however, had it been a constant cross-section tube, this time would have been 570$\mu$s. Therefore, this clearly indicates that inside the converging section, the incident shock is accelerating.
	
		\begin{figure}[t]
		\centering
		\includegraphics[width=0.8\linewidth]{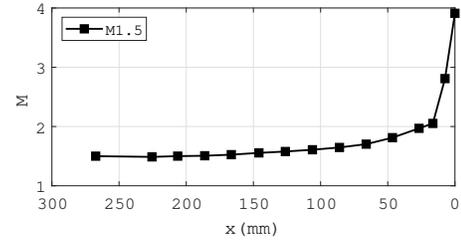}
		\caption{Mach number variation within the converging section along the central axis for the case M1.5}
		\label{MM1p5}
	\end{figure}
	
	In view of this, speed of the shock while moving through the converging section is estimated to realize the influence of focusing phenomenon on the shock strength. The static pressure profile at various points along the axis inside the converging section is monitored from which the shock Mach number is calculated through locating the time instances of incident shock induced pressure rise at those points. The initial shock strength of Mach 1.5 gets accelerated to 3.9 Mach as it reaches the focusing point which can be perceived from Figure ~\ref{MM1p5}. This is in line with the fact that SW accelerates as it encounters a reduction in the cross-sectional area \cite{russell1967}. Nevertheless, major component of the shock acceleration happens to be inside conical insert present at the end of the converging section, where the incident shock strength varies exponentially.
	
	\begin{figure}[t]
		\centering
		\includegraphics[width=1\linewidth]{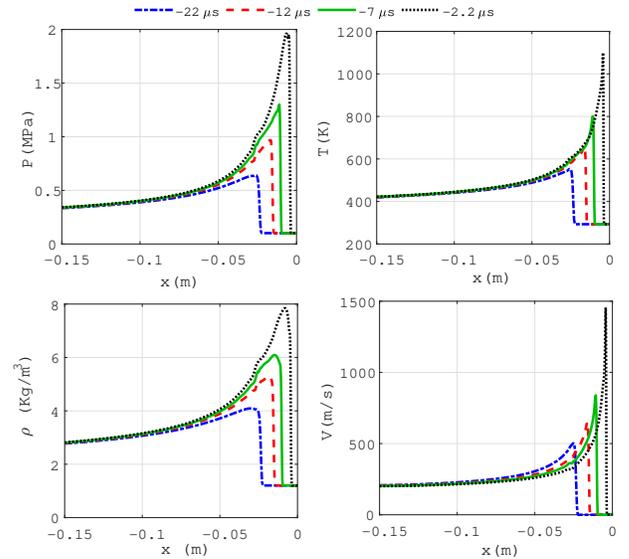}
		\caption{Flow properties like pressure, temperature, density and axial velocity monitored along the central axis of the domain for case M1.5 at four different time instances before focusing}
		\label{axisline}
	\end{figure}
	
	\begin{figure}
	\centering
	\includegraphics[width=0.9\linewidth]{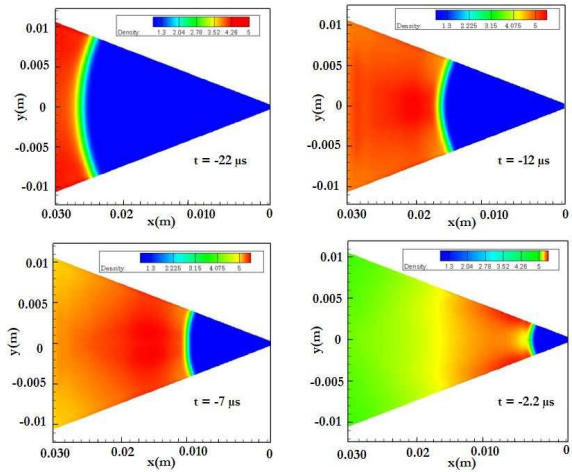}
	\caption{Instantaneous density contour at various instances from a time interval of 22 $\mu$s time to 2.2 $\mu$s before the shock reaches the focusing point}
	\label{conedens}
\end{figure}	
	
	In perspective of significant shock Mach number variation, fluid properties at different time instances along the axis in the near focus region is quantified as shown in Figure ~\ref{axisline} . The associated location and shape of incident shock for these time stamps can be depicted from Figure ~\ref{conedens}. Drastic increment of all the flow parameters is observed as shock approaches the focusing point. This can be merely due to the associated energy of SW getting concentrated into a very small region. Moreover, exponential increment of shock strength is also a potential factor in enhancing the fluid property magnitudes. At the locus of focusing point, the static pressure and temperature values are recorded to be 115 MPa and 5600 K respectively for the initial filling conditions corresponding to Case-1.

\begin{figure}[h!]
	\centering
	\includegraphics[width=0.9\linewidth]{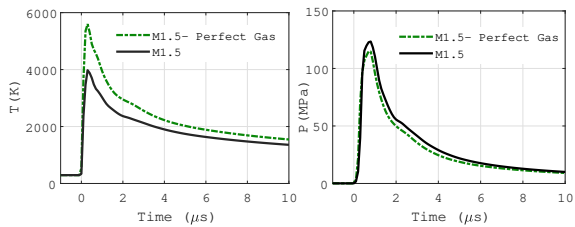}
	\caption{Peak values comparison of temperature and pressure obtained at focusing point of case M1.5 for simulations with and without high-temperature effects}
	\label{TP-Perf}
\end{figure}

	As the perfect gas simulation resulted in significantly high magnitude of pressure and temperature at the focusing point, it can thereby trigger dissociation of the test gas viz. air. Therefore, simulations incorporating high temperature effects are carried out where different chemical reactions and temperature dependent fluid property models are taken into account. Comparative assessment of perfect gas and high temperature simulations with the obtained peak values of pressure and temperature is shown in Figure ~\ref{TP-Perf}. 
	
	After including high temperature effects to the simulation, the decrement in focusing point temperature is prominent whereas less variation in peak pressure is observed. The peak temperature is found to be reduced from 5600 K to 4000 K while 7\% increment in peak pressure is noted. Energy utilization in rotational as well as vibrational excitation and dissociation of air molecules would have resulted in the decrement of peak temperature. Nevertheless, as the compressibility factor of air approaches one under high temperature and high pressure circumstances; hence, the peak value of pressure might have increased with peak temperature dropping. With this insight of high temperature effect, it is thereby planned to alter the filling conditions so as to quantify the flow parameters and species concentration.
	
	\subsection{\label{sec:level3.1}Effect of initial fill pressure}
	
\begin{figure}[h!]
		\centering
		\includegraphics[width=0.9\linewidth]{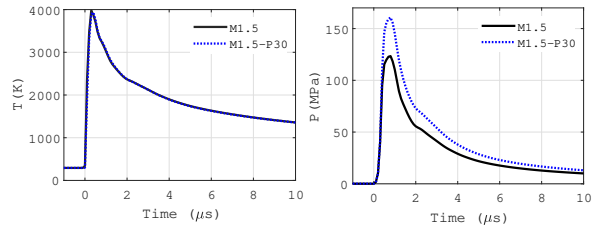}
		\caption{Peak values comparison of temperature and pressure point showing the effect of initial fill pressure of the test gas obtained at focusing when the shock strength is maintained constant}
		\label{TP-P}
	\end{figure}

	The initial fill pressure of the test gas is increased by 30\% (Case-2) to analyze the effect of the fill pressure on the shock focusing phenomenon. The static temperature and pressure monitored throughout the simulation at the focusing point for the base case along with Case-2 are shown in Figure ~\ref{TP-P}. The peak pressure magnitude is calculated to have an increment of 30\% as compared to the base case which seems to exhibit as an immediate consequence due to change in filling pressure. However, the peak temperature for Case-2 shows a 2\% reduction as compared to Case-1 although the fill temperature is kept constant. The decrement in temperature is certain, as this effect along with increment in peak pressure would lead towards unity compressibility factor. As far as trend of flow parameters are concerned, eventually for both the cases, the temperature and pressure profiles reach the equilibrium value at $~$10 $\mu$s after focusing.

	The resulted temperature and pressure at the focusing point turns out to be high enough to activate the dissociation of the test gas. However, the increased magnitude holds only for a few microseconds before attaining the equilibrium values. The mass fraction concentrations of all the nine species including ions are monitored for both the test cases during this duration at the focusing point.
	
			\begin{figure}
		\centering
		\includegraphics[width=0.9\linewidth]{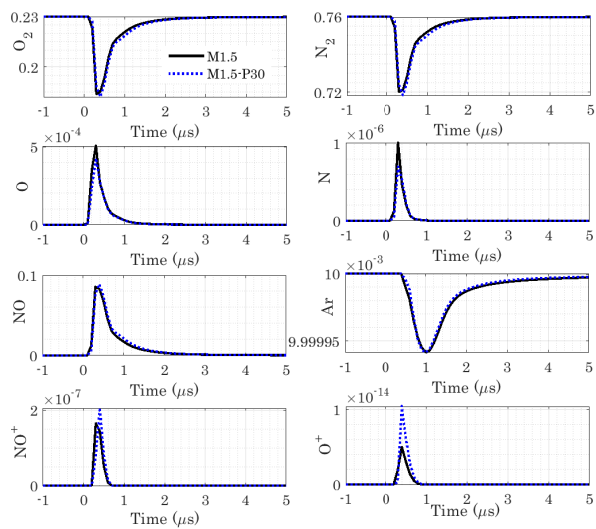}
		\caption{Species mass fraction variation showing the effect of initial fill pressure of the test gas obtained at focusing point when the shock strength is maintained constant}
		\label{speciesP}
	\end{figure}
	
	The effect of initial fill pressure on the focusing phenomenon is presented in terms of variation in species concentration in Figure ~\ref{speciesP}. The dissociation of $N_2$ and $O_2$ produces finite amount of atomic nitrogen (N), atomic oxygen (O) and Nitric oxide (NO). After achieving the peak magnitude, these species viz. `N', `O' and NO further recombines to help in the formation of $N_2$ and $O_2$. This is evident as the concentration of `N', `O', and NO continuously decay after attaining a peak value, whereas mass fractions of $N_2$ and $O_2$ are increasing. After few microsecond, as species `N' and `O' vanishes; thereafter,$N_2$, $O_2$ and NO interact among each other through third body reactions till the attainment of respective equilibrium values.
	
	As far as species concentration for both the cases are concerned, the dissociation of $N_2$ and $O_2$ is observed to have increased moderately for M1.5-P30. The key reason for higher dissociation could be the increased number of molecule interactions due to the peak pressure increment. The peak values of `N' and `O' has reduced with higher decrement for species `N'. However, this reduction is balanced with increased magnitude of NO species which enforces the possible recombination of `N' and `O' to form NO. The trend of attaining equilibrium for all the species is similar in both cases. It is observed that species `N' and `O' attains equilibrium faster compared to $N_2$, $O_2$ and NO. The longer time taken by species NO to reach equilibrium in turn extends the non-equilibrium duration for $N_2$ and $O_2$. Argon takes the longest duration to reach equilibrium compared to all other species. 
	
	Another observation from the Figure ~\ref{speciesP} is that the initial pressure variation in the driven section is largely affecting the ionization reactions. Noticeable increment in the peak is observed for the ions. However, the magnitude of mass fraction of ions is found to be very less with Ar+ ion being the lowest which is in the order of $10\textsuperscript{-16}$. The lower concentration of ions is due to unavailability of enough activation energy to trigger the ionization reactions significantly. Nevertheless, after attaining equilibrium, the mass fraction for all species is found to be same for both the cases which is also equal to the initial concentration of the test gas. Furthermore, the duration of the dissociation and ionization reaction is mostly dependent upon the activation energy magnitude. The value of the activation energy will be higher during first few microseconds after focusing, and as time elapses, the magnitude will decrease. However, the activation energy requirement for ionization is higher followed by dissociation and recombination reaction. Therefore, as can be seen from Figure ~\ref{speciesP}, the ionization reaction gets over first and followed by the dissociation reaction.

	\subsection{\label{sec:level3.2}Effect of initial fill temperature}
	
\begin{figure}[h!]
	\centering
	\includegraphics[width=0.9\linewidth]{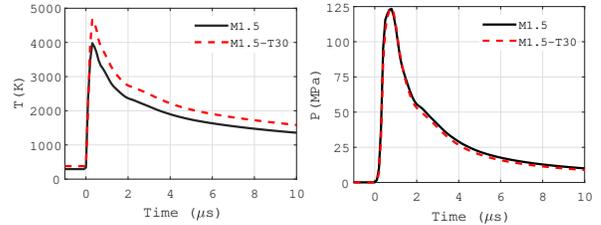}
	\caption{Peak values comparison of temperature and pressure point showing the effect of initial fill temperature of the test gas obtained at focusing when the shock strength is maintained constant}
	\label{TP-T}
\end{figure}

	The study of initial fill temperature effect is obtained by varying the driven gas temperature by 30\% (Case-3). The maximum temperature and pressure obtained at the focusing point due to this alteration is shown in Figure ~\ref{TP-T}. The temperature peak value has increased by 18\% for Case-3 as compared to that of the base case viz. Case-1. After attaining equilibrium, still the value of temperature remains high by 14\% for M1.5-T30. The static pressure at the focusing point for M1.5-T30 seems to be unaltered with respect to M1.5.

	During the previous case study, it can be seen that initial pressure increment of 30\% resulted in 30\% increase in peak pressure at the focusing point. But in the current case, initial temperature rise of 30\% yields in 18\% enhancement in the peak temperature. This is mostly because of the high temperature which resulted in higher reaction rate for the species thereby utilizing some of the flow energy in excitation of the molecules. As a result, effective temperature increment becomes less.
	
\begin{figure}[h!]
	\centering
	\includegraphics[width=0.9\linewidth]{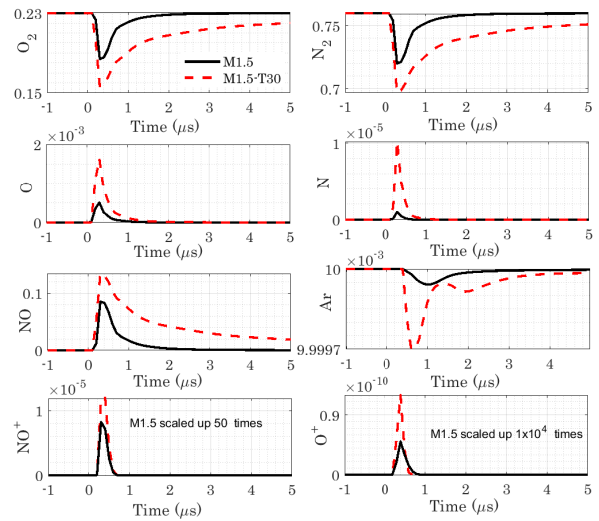}
	\caption{Species mass fraction variation showing the effect of initial fill temperature of the test gas obtained at focusing point when the shock strength is maintained constant. The magnitude of $NO^+$ and $O^+$ ions of M1.5 is magnified in order to have a comparative assessment.}
	\label{speciesT}
\end{figure}
	
	The effect of initial fill temperature on species concentration is shown in Figure ~\ref{speciesT}. The fill temperature is found to have much more influence on the focusing phenomenon than fill pressure. Due to higher temperature, higher number of molecular collisions per unit time will take place, which thereby increases the reaction rate and further the species formation. $N_2$ and $O_2$ dissociation has increased considerably and therefore higher mass fraction values for `N', `O' and NO. In contrary to the observation in Figure ~\ref{speciesP}, here the percentage increment in `N' and `O' species seems to be higher than that of NO species prediction; therefore, indicating towards more activation of endothermic dissociation reactions with increased fill temperature.
	
	The higher equilibrium temperature for M1.5-T30 further leads to elevated magnitude of equilibrium species concentration for $N_2$, $O_2$ and NO species. Simultaneously, argon is having third body reaction with oxygen till it reaches equilibrium value. As, the ionization reactions are dependent on both pressure and temperature \cite{fruth1923}; therefore, significant increment is also perceived for $NO^+$ and $O^+$ ions due to higher fill temperature. But, upon observation from Figure ~\ref{speciesP} and Figure ~\ref{speciesT}, it is clear that the dependency of ion concentration on fill temperature is much higher as compared to the fill pressure. Also, the maximum mass fraction achieved by $Ar^+$ ion is found to be increased to a range of $10^-12$. Consequently, the initial fill temperature seems to have a notable effect on the magnitude of species formation as compared to the fill pressure.

\subsection{\label{sec:level3.3}Effect of initial Shock strength}

\begin{figure}[h!]
	\centering
	\includegraphics[width=1\linewidth]{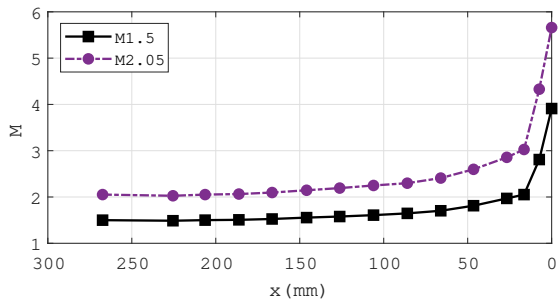}
	\caption{Mach number variation across the central axis of the converging section showing effect of initial shock strength}
	\label{mm}
\end{figure}

	Finally, the effect of shock strength on the shock focusing phenomenon is studied and results are discussed here. The shock acceleration within the converging section for both the cases is depicted in Figure ~\ref{mm}. The initial percentage change of 36\% in the shock strength increases to 42\% at the focus point. As discussed earlier, the major acceleration and strengthening of the shock seems to be happening inside the conical section which is close to the focusing point.

	\begin{figure}
		\centering
		\includegraphics[width=1\linewidth]{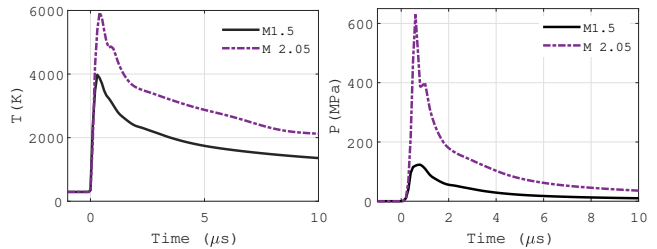}
		\caption {Peak values comparison of temperature and pressure showing the effect of initial shock strength obtained at focusing point }
		\label{tp-m}
	\end{figure}

	The comparison of the static temperature and static pressure monitored at the focusing point is shown in Figure ~\ref{tp-m}. This reveals that with increased shock strength, the peak temperature increased by 50\% and also the pressure magnitude is having huge increment. This is also in line with the fact that due to higher Mach numbers, the pressure ratio will have significant increment as compared to the temperature ratio \cite{anderson}. It is also noticeable from the figure that time taken by the flow to reach equilibrium state is higher for increased shock strength. Besides, after also reaching equilibrium, the equilibrium pressure as well as temperature for M2.05 remains higher than that of M1.5.
	
		\begin{figure}[h!]
		\centering
		\includegraphics[width=0.9\linewidth]{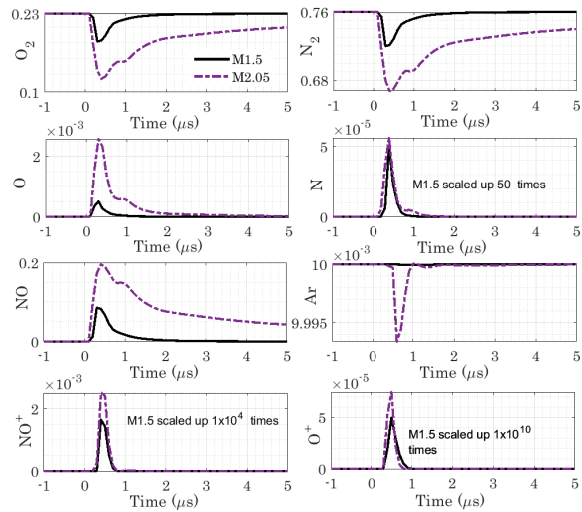}
		\caption{Species mass fraction variation showing the effect of initial fill pressure of the test gas obtained at focusing point when the shock strength is maintained constant}
		\label{speciesM}
	\end{figure}

	The mass fraction of species monitored at the focusing point showing the effect of the shock strength is depicted in Figure ~\ref{speciesM}. With increase in shock strength, the dissociation rate of $N_2$ and $O_2$  is increased, and the formation of `N', `O', NO and ions has peaked up. The increment in species concentration is the cumulative effect of increment in peak pressure and temperature. In line with this, significant increase in species production is observed for all the species with increase in increase in species production is observed for all the species with increase in shock strength. However, the increment of species `N' is less as compared to the others as opposing effect is expected to happen independently for pressure and temperature rise. For the ions, the difference in peak value is large enough that the distribution of M1.5 looks like a straight line compared to that of M2.05. This shows that the ionization reaction rate is increased considerably. The delay in attaining equilibrium for $N_2$, $O_2$ and NO is also observed.
	
	\section{\label{sec:level4}Conclusion}
	
	A detailed study on the phenomenon of shock wave focusing is carried out numerically. Spherical shock wave focusing is achieved with the help of a perfectly contoured converging section attached to a shock tube. Computations with inviscid perfect gas effects as well as with high temperature effects are carried out with air as test gas. Dissociation, recombination and ionization reaction comprising of nine prominent species including ions ($N_2$, $O_2$, N, O, NO, Ar, $NO^+$, $O^+$ and $Ar^+$) are incorporated. Also, consequence of high temperature effect is assessed through comparative assessment of the peak temperature and pressure obtained at the focusing point. The effect of initial fill condition and initial shock strength on the focusing phenomenon is also studied. During assessing the effect of initial fill condition, fill temperature and fill pressure is independently increased by 30\%. Nevertheless, the effect of altering the filling temperature seems to have noteworthy effect on focused regime.
	
	As direct consequence of change in fill pressure, the peak pressure at the focusing point is increased by 30\% while the peak temperature reduces marginally with increased value of NO. However, decrement in species `N' and `O' is perceived. With 30\% increase in the fill temperature, the peak pressure remains same whereas peak temperature at the focusing point is increased by 18\% as the remaining energy is utilized in excitation of the molecules. The enhanced number of molecular collision has resulted significant increment in all the species concentration. The dependence of ionization rate is also found to be much more receptive towards the fill temperature rather than fill pressure. Finally, the effect of shock strength is evaluated which indicates increment in both peak temperature as well as peak pressure due to higher shock Mach number in the near focus region. Moreover, this resulted in analyzing combining effect of the aforementioned variations. This in turn influence the species concentration where huge variation in all species concentration is obtained except for `N' due to conflicting effect is expected to happen independently for pressure and temperature rise. All the ion formation is also seen to have towering increment. 
	
	\begin{acknowledgments}
		The work was supported by Department of Science and Technology (DST), India, under the Early Career Research Award, ECRA/2018/000678.
	\end{acknowledgments}

	
	%

\end{document}